# Fast continuous ion transport through X-junctions with low motional excitation for QCCD architectures

**Jie Xiao[1,2,†], Zhao Wang [3,†,*], Chenghong Zhu[4], Shiqiang Gao[1,2], Xin Wang[5,*], Fa Dai[3,*], Xu Wang[1,*]**

[1]Key Laboratory of Advanced Manufacturing Technology of the Ministry of Education, Guizhou University, Guiyang 550025, China

[2] MoonQuest AI Fundamental Technology Research (Shenzhen) Co., Ltd., Guangzhou 510006, China

[3] School of Integrated Circuits, Guangdong University of Technology, Guangzhou 510006, China

[4]QudeLeap Research, Shanghai 200030, China

[5]Thrust of Artificial Intelligence, Information Hub, The Hong Kong University of Science and Technology (Guangzhou), Guangzhou 511453, China

†These authors contributed equally to this work.

*E-mails: joeshardow@gmail.com (Zhao Wang), felixxinwang@hkust-gz.edu.cn (Xin Wang), daifa@gdut.edu.cn (Fa Dai), xuwang@gzu.edu.cn (Xu Wang)

Trapped ion quantum computing based on the quantum charge coupled device (QCCD) architecture uses functional zoning and ion transport to enable high fidelity computation and all to all qubit connectivity. As trapped ion quantum processing units (QPU) scale, 90° X-junction turns account for an increasing fraction of total transport time. Frequent junction crossings, therefore, make transport speed and final phonon excitation direct determinants of quantum algorithm execution efficiency. However, complex electric fields in X-junctions complicate waveform design, making it difficult to achieve high speed and low final phonon excitation simultaneously. Here, we develop a waveform design method for low-phonon-excitation high-speed X-junction transport (LHX-T). It integrates low barrier pseudopotential channel design using multiple radio frequency (Multi-RF), multidimensional shortcut to adiabaticity (STA) dynamics, and constrained DC voltage optimization over the entire transport sequence. This framework connects waveform design, control, and computational validation, with smooth multichannel waveforms computed rapidly through a Karush–Kuhn–Tucker (KKT) linear system. The resulting waveform transports a single $^{171}Yb^{+}$ ion through a 90° turn in 5 $\mu s$ at an average speed of 14.6 $m/s$. Over 50 consecutive round trips, the final equivalent coherent phonon number consistently stays below 0.001, with no noticeable accumulation over time. System level evaluations at the QCCD scale show that the median duration of a single execution cycle is reduced by about 30% across various circuit types. The fraction of execution time spent on transport decreases from approximately 70% to 55%. This approach offers a means of designing fast ion transport waveforms with low phonon excitation for scalable QCCD based QPUs and provides a technical basis for efficient execution of quantum algorithms.

Long coherence times[1,2], exceptionally high fidelity quantum operations[3,4], and all to all connectivity[3,5] make trapped ion systems an important experimental platform for practical quantum computing. As ion numbers increase, conventional one dimensional linear traps encounter several bottlenecks. These include reduced trapping stability[6], non-uniform ion spacing[7], spectral crowding of phonon modes[1,5,8], and increasing complexity in individual ion addressing[1,5]. Chip based trapped ion quantum computing with a QCCD architecture[4,9,10] addresses these limitations through dedicated storage, operation, and detection zones. Ion transport enables global scheduling and reordering, substantially reducing noise crosstalk and improving operational fidelity while retaining all to all connectivity[3,4,11]. Previous studies report that transport accounts for 70% of a circuit execution cycle when quantum algorithms are implemented on one dimensional ion chains[3,4,12]. As more ions are controlled simultaneously on a chip, X-junction transport occupies an increasing fraction of the total transport time[13–17]. Faster X-junction transport with lower final phonon excitation can therefore accelerate circuit execution and shorten execution cycles[18,19], providing an important basis for scaling trapped ion QPU.

Existing 90° X-junction transport schemes fall into two categories according to their paths: turns along segmented straight paths[13–15,17,20] and continuous curved paths[16,21]. Segmented schemes[14,15,17] typically slow and pause the ion near the junction center, while control voltages rotate the local principal axes. This achieves low excitation turning, but the intermediate pause increases the total junction crossing time. Continuous curved path schemes[16,21] avoid this pause, but transport remains slow. Their waveform design neither incorporates nor constrains final phonon excitation, and corresponding experimental measurements have not been reported. Multidimensional STA methods[22] were developed to increase transport speed while maintaining low excitation. For turning transport, however, they specify a spatial potential that evolves with time without providing its decomposition into experimentally applicable voltage waveforms. Methods exist for directly decomposing the entire target potential into time dependent electrode contributions[21,23]. Our simulations and tests nevertheless indicate that low excitation STA transport requires highly accurate potential reconstruction and a large number of electrodes for fitting. Otherwise, substantial final phonon excitation can arise. Specifying a geometric path and a fixed trap shape alone, therefore, does not ensure low final excitation during fast multidimensional STA turning transport. The outcome also depends on the forces acting on the ion, local confinement, and mode evolution[22]. These dynamical targets must ultimately be mapped to smooth, continuous control waveforms using a finite number of electrode degrees of freedom. Pseudopotential barriers, weakened confinement, and rapid changes in mode orientation near X-junctions further complicate this mapping[14–16,24].

A recently proposed cooperative optimization method using multiple RF (Multi-RF)[24] adjusts segmented RF electrode voltages to shape a single connected pseudopotential channel for continuous turning. The channel combines a low X-junction barrier with small variations in ion height and trap frequency, while the principal axis orientations change continuously. These properties make it well suited to continuous turning transport. Building on this channel, we develop LHX-T for continuous ion turning without intermediate stops and with low phonon excitation. The method integrates Multi-RF[24] confinement optimization, multidimensional STA dynamics[22], and constrained DC voltage optimization over time into a complete waveform design and control framework. Unlike approaches that map the entire target trap potential onto electrode voltages[21,23], LHX-T imposes the target force along the STA trajectory as an explicit equality constraint. Local curvature is incorporated into the quadratic optimization objective, substantially improving the accuracy of force reproduction and waveform computation. Molecular dynamics simulations validate the computed transport waveforms. The local mode evolution and final coherent motional excitation after 50 round trips agree closely with their design targets, with low excitation and no accumulation. Finally, using the S-SYNC assessment framework[25], we analyze how the total X-junction transport time and total ion transport time affect quantum algorithm execution time as QCCD chips scale (Fig. 5). The frequency and relative contribution of X-junction turns increase substantially with chip size. Faster turns, therefore, reduce the fraction of the algorithm's execution time spent on transport, thereby improving execution efficiency. Across the quantum algorithms considered, the median reduction per computation cycle is approximately 30%. For some algorithms, execution time decreases by approximately 40%, at which point transport is no longer the dominant bottleneck.

## Results

### LHX-T control and validation framework

The Multi-RF method forms a single connected pseudopotential channel near the X-junction that supports continuous turning (Fig. 1). Using the Lewis–Riesenfeld invariant and its multidimensional STA extensions[22,26–28], we construct the target trajectory, force, local curvature, and mode orientation conditions required by LHX-T. We explicitly constrain the STA target force, force balance along the $z$ axis, and minimize total potential Hessian coupling between the $xy$ plane and $z$. Within the feasible set, STA curvature matching, voltage amplitudes, and temporal smoothness are optimized. This strategy reconciles competing STA objectives given the limited degrees of freedom available for the DC electrodes. All time points are incorporated into a single quadratic optimization problem over the entire transport sequence. Lagrange multipliers yield a KKT linear system, accelerating the computation of smooth multichannel DC voltage waveforms for fast transport. The detailed derivation is provided in the Methods section. To validate the waveforms, we apply the optimized DC voltages to the corresponding electrode models. Their potentials are added to the RF pseudopotential to obtain the realized three dimensional time dependent total potential, $\Phi_{\mathrm{tot}}(\mathbf{r},t)$. We then simulate the forces, local trap shape, and ion position over time and compare them with the STA targets (Fig. 2 and Fig. 3). The close agreement supports the validity and effectiveness of the generated waveforms within the simulation framework. Finally, phonon excitation simulations over 50 consecutive round trips tests the low excitation performance (Fig. 4). The low excitation supports the stability of the computed transport waveforms.

### X-junction and pseudopotential channel

We use the two dimensional surface electrode X-junction with segmented RF electrodes reported in Ref. 24 as the validation geometry (Fig. 1). It comprises four transport regions, A–D; Fig. 1 also shows the transport DC electrodes, central electrode M1, and ground electrodes. Cooperative optimization of the segmented RF amplitudes over 150–250 V yields a single connected, continuous, low barrier pseudopotential channel for a 90° turn at a height of 52 $\mu m$. The red dashed line in Fig. 1 traces the RF null in three dimensions. The translucent green region is the 0.8 $mV$ pseudopotential isosurface, illustrating the continuous channel. The black dashed line shows the projection of the prescribed 90° turning trajectory onto the $xy$ plane. The trajectory lies at a height of 52 $\mu m$. It comprises a quarter-circular arc of radius 40 µm and two straight segments, each 5 µm long, one at either end.

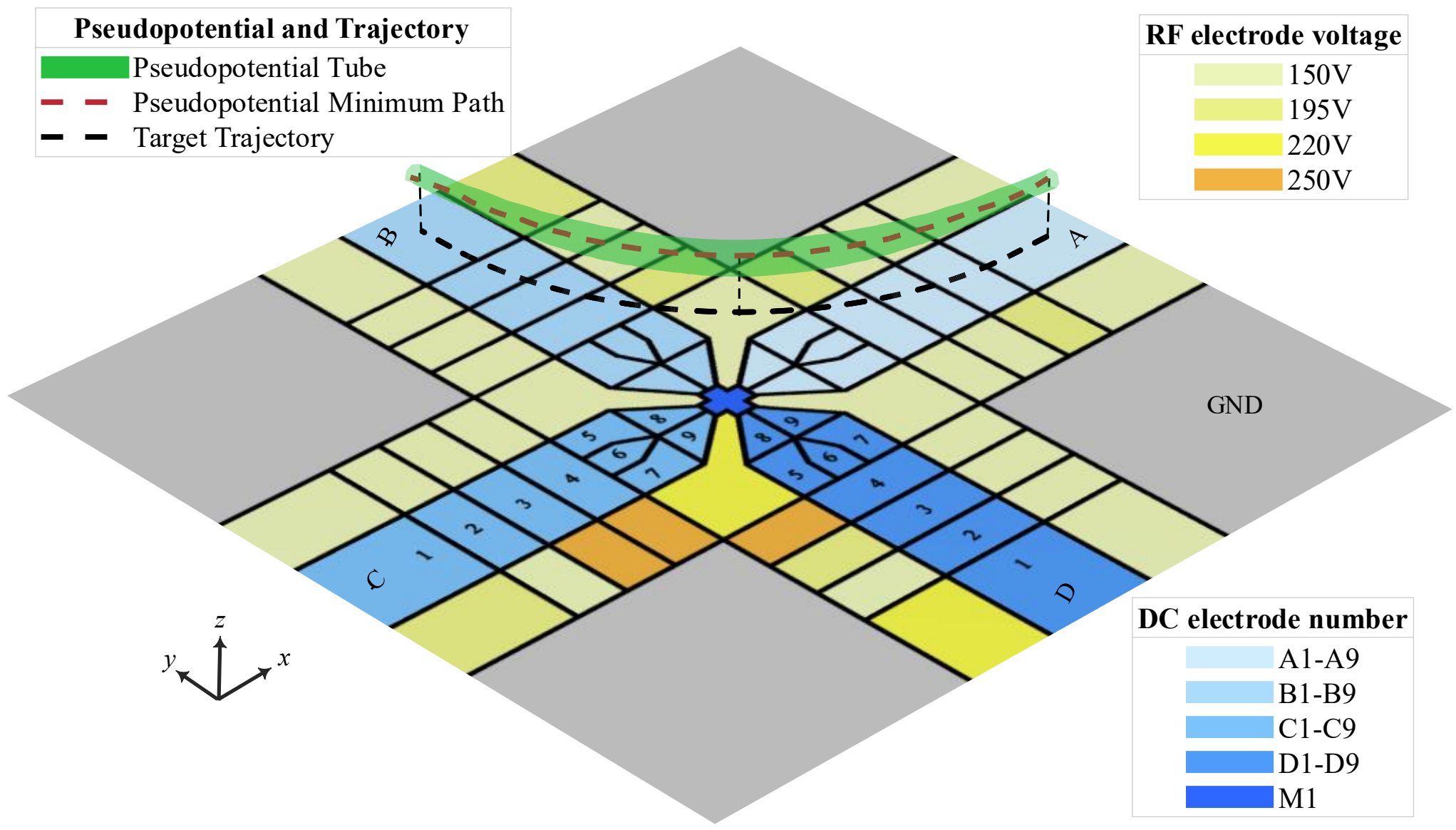

**Fig. 1 | 2D surface electrode X-junction, Multi-RF pseudopotential channel, and target turning trajectory.** RF electrodes are shown in yellow, with shades indicating amplitudes of 150, 195, 220, and 250 V. DC electrodes are shown in blue. The four transport arms are designated A–D, each containing nine electrodes numbered 1–9 in a rotationally symmetric arrangement. The central electrode M1 is dark blue, and the remaining grey regions are grounded electrodes. The translucent green region shows the pseudopotential channel formed through cooperative optimization of the Multi-RF[24]. The red dashed line traces the RF null in three dimensions, and the black dashed line shows the planar projection of the target turning trajectory.

### DC voltage optimization and physical constraints

Optimization over the entire 5 $\mu s$ transport interval yielded smooth, continuous DC voltage waveforms ranging from 0 to 15 V (Fig. 2a–d). Fig. 2e compares the target and realized forces $F_{x,y,z}$ acting on the ion along the target trajectory in $\Phi_{\mathrm{tot}}(\mathbf{r},t)$. The components $F_x$ and $F_y$ provide the acceleration, deceleration, and centripetal force required for planar turning, $F_z$ remain near zero to maintain the transport height. The realized forces closely followed their targets. Fig. 2f compares the target and realized Hessian components $M_{xx}$, $M_{yy}$ and $M_{xy}$ of $\Phi_{\mathrm{tot}}(\mathbf{r},t)$ at the target trajectory. The components $M_{xx}$ and $M_{yy}$ describe the local curvatures along $x$ and $y$, respectively, while $M_{xy}$ describes their cross curvature. Together, these components determine the local in-plane mode orientations[22,29–31]. The realized curvatures broadly followed their targets. Thus, a single set of smooth DC waveforms reproduced both the target forces and local trap curvatures.

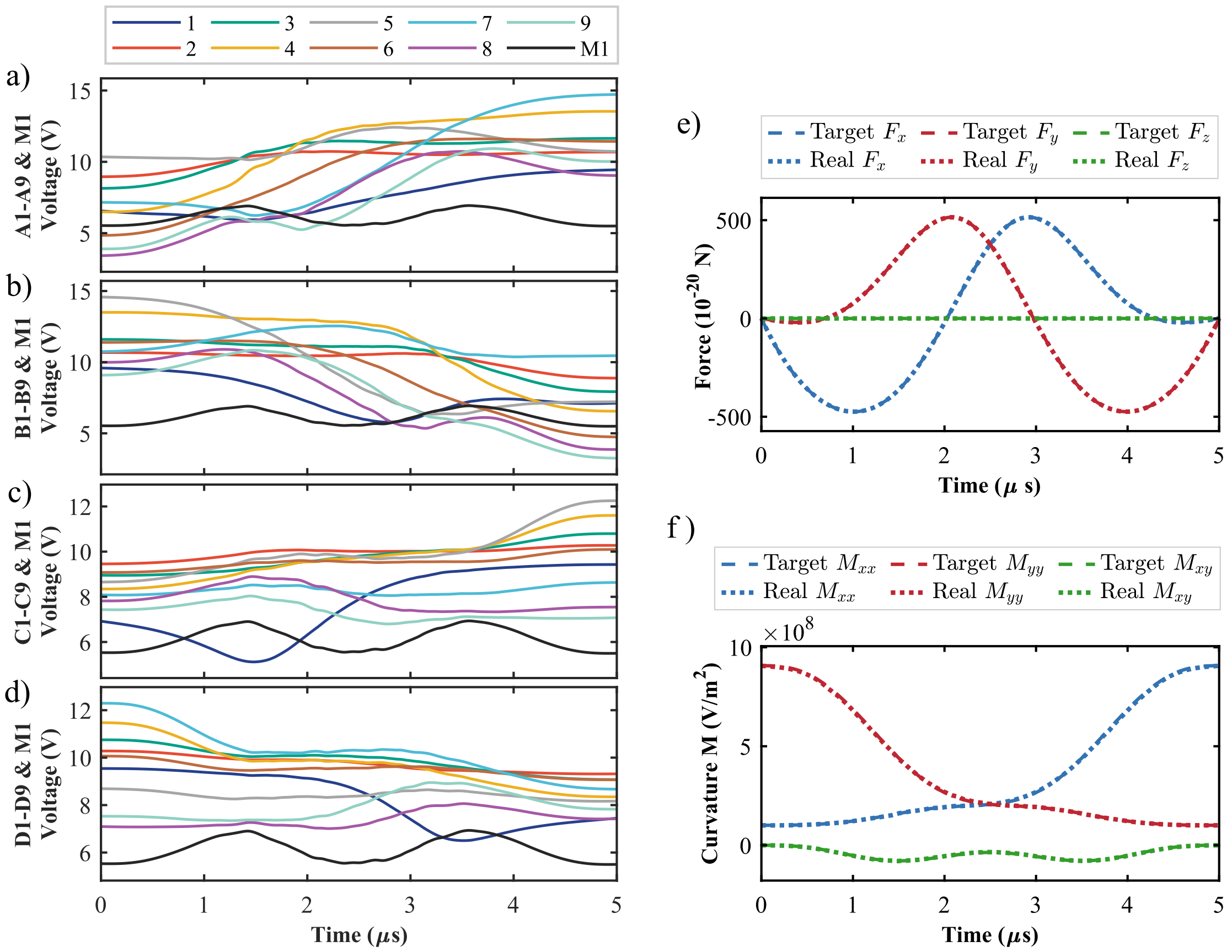

**Fig. 2 | Multichannel DC waveforms optimized over the entire transport sequence and realization of the constraint targets. a–d**, Voltage waveforms for DC electrodes 1–9 in transport arms A–D and the central electrode M1. Matching numbers identify rotationally equivalent electrodes in different arms; black curves denote M1. **e**, Target (dashed curves) and realized (dotted curves) x, y, and z force components at the target trajectory. **f**, Target (dashed curves) and realized (dotted curves) local trap curvature components $M_{xx}$, $M_{yy}$ and $M_{xy}$.

### Principal axis orientations and phonon modes

At each time point, we locate the local minimum $C_{x,y,z}$ of $\Phi_{\text{tot}}(\mathbf{r},t)$, which defines the trap center. Diagonalizing the Hessian at this position[31] yields the realized mode frequencies and orientations. Fig. 3a compares the target and realized positions $C_{x,y,z}$, which broadly agree throughout the transport. The largest deviations occur along the weakly confined direction near the beginning and end of the sequence. Its smaller curvature makes $C_{x,y,z}$ more sensitive to potential deviations[32,33].

During continuous two dimensional turning, small mode frequency separations increase the sensitivity of mode orientations to local potential perturbations[34] and the likelihood of non-adiabatic mode coupling[35]. Fig. 3b shows the evolution of the local trap frequencies. Initially, the axial mode $\omega_a$ and radial mode $\omega_r$ in the $xy$ plane have frequencies of 1.2 MHz and 3.6 MHz, respectively. Their frequencies approach one another during the turn, reaching a minimum separation of approximately 0.35 MHz near the trajectory midpoint. They subsequently separate and return to their initial values at the endpoint. Because the DC electrostatic potential satisfies Laplace's equation[36], the $z$ curvature is not an independent control variable, and no separate target $\omega_z$ is imposed. The realized $\omega_z$ is jointly determined by the RF pseudopotential and DC potential. It remains above $\omega_a$ and $\omega_r$ throughout transport and returns to its initial value at the endpoint.

Fig. 3c shows the evolution of local mode orientations. The angles $\theta_a$, $\theta_r$ and $\theta_z$ are defined between the Hessian eigenvectors[31] and the positive $x$, $y$ and $z$ reference axes, respectively. During transport, $\theta_a$ changes continuously from 0° to 90°, while $\theta_r$ follows the opposite trend. The modes $\omega_a$ and $\omega_r$ exchange orientations during the 90° turn, whereas $\theta_z$ remains near 0°. The realized orientations broadly follow their targets without abrupt changes near the minimum frequency separation, maintaining continuous evolution of non-degenerate local modes.

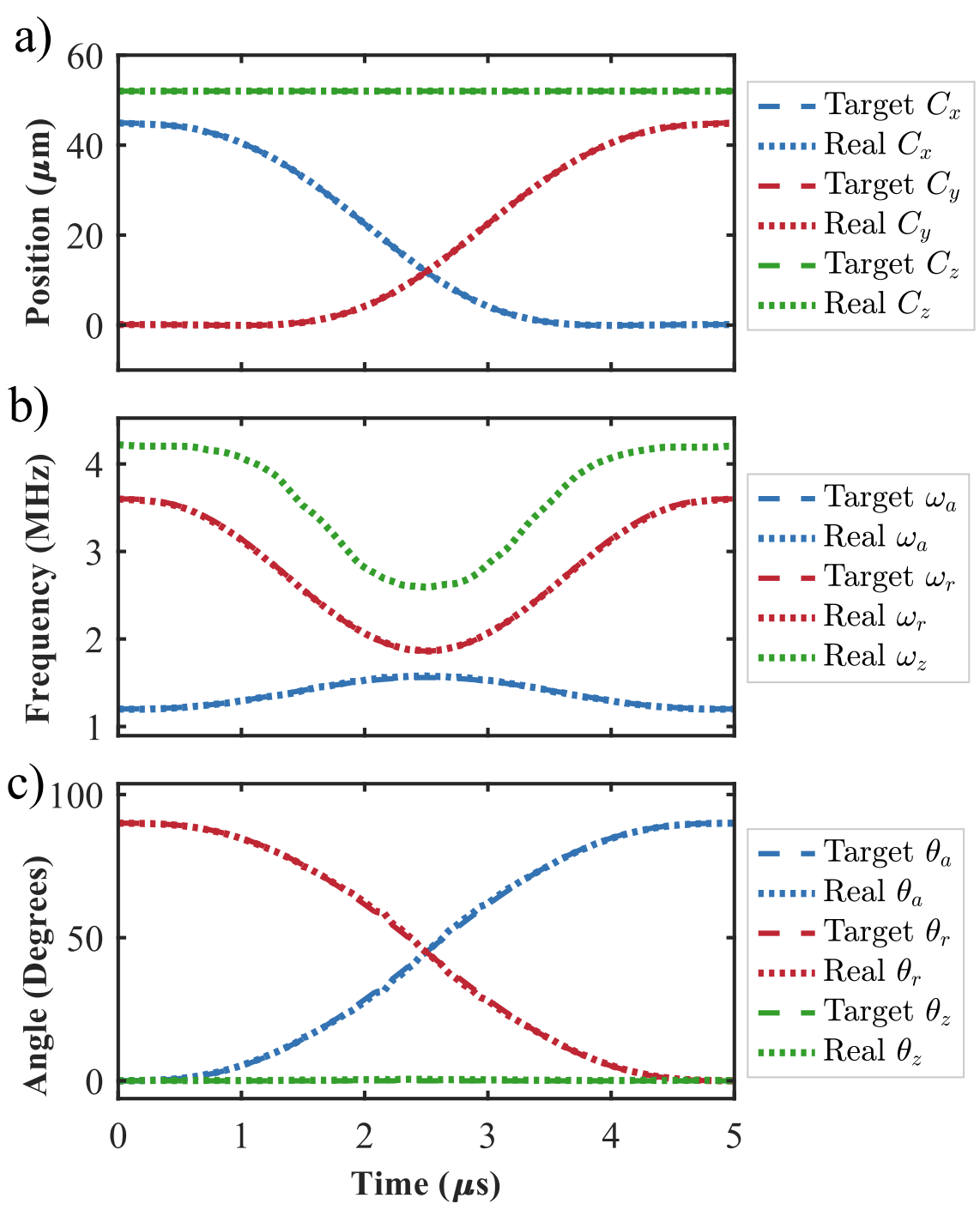

**Fig. 3 | Local trap centers, mode frequencies, and orientations during turning transport. a**, Target (dashed curves) and realized (dotted curves) trap center position $C_{x,y,z}$. **b**, Target (dashed curves) and realized (dotted curves) frequencies of the two eigenmodes $\omega_a$ and $\omega_r$ in the $xy$ plane, together with the realized z mode frequency $\omega_z$. **c**, Target (dashed curves) and realized (dotted curves) mode orientations. The angles $\theta_a$, $\theta_r$ and $\theta_z$ are defined between the directions of modes $\omega_a$, $\omega_r$ and $\omega_z$ and the positive $x$, $y$ and $z$ reference axes, respectively.

**Dynamical validation and final coherent motional excitation**

Residual displacement and velocity after fast transport produce final coherent motional excitation[37,38] and can introduce quantum gate errors[30]. We simulate ion dynamics in the realized three dimensional, time-dependent total potential $\Phi_{\text{tot}}(\mathbf{r},t)$. The target and simulated trajectories nearly coincide throughout the 90° turn completed in 5 $\mu s$ (Fig. 4a). Fig. 4b shows the phase-space evolution over one round trip. The trajectories along $x$ and $y$ return to their initial phase-space points, while position and velocity along $z$ remain nearly unchanged. These results indicate very small residual coherent motion at the endpoint. Fig. 4c shows the final phonon excitation over 50 consecutive round trips. Excitation $n_a$ of the low frequency mode $\omega_a$ exhibits periodic oscillations on the $10^{-6}$ phonon scale. Excitations $n_r$ and $n_z$ of the higher frequency modes $\omega_r$ and $\omega_z$ remain close to zero. None of the three modes shows sustained growth in excitation as the number of round trip increases.

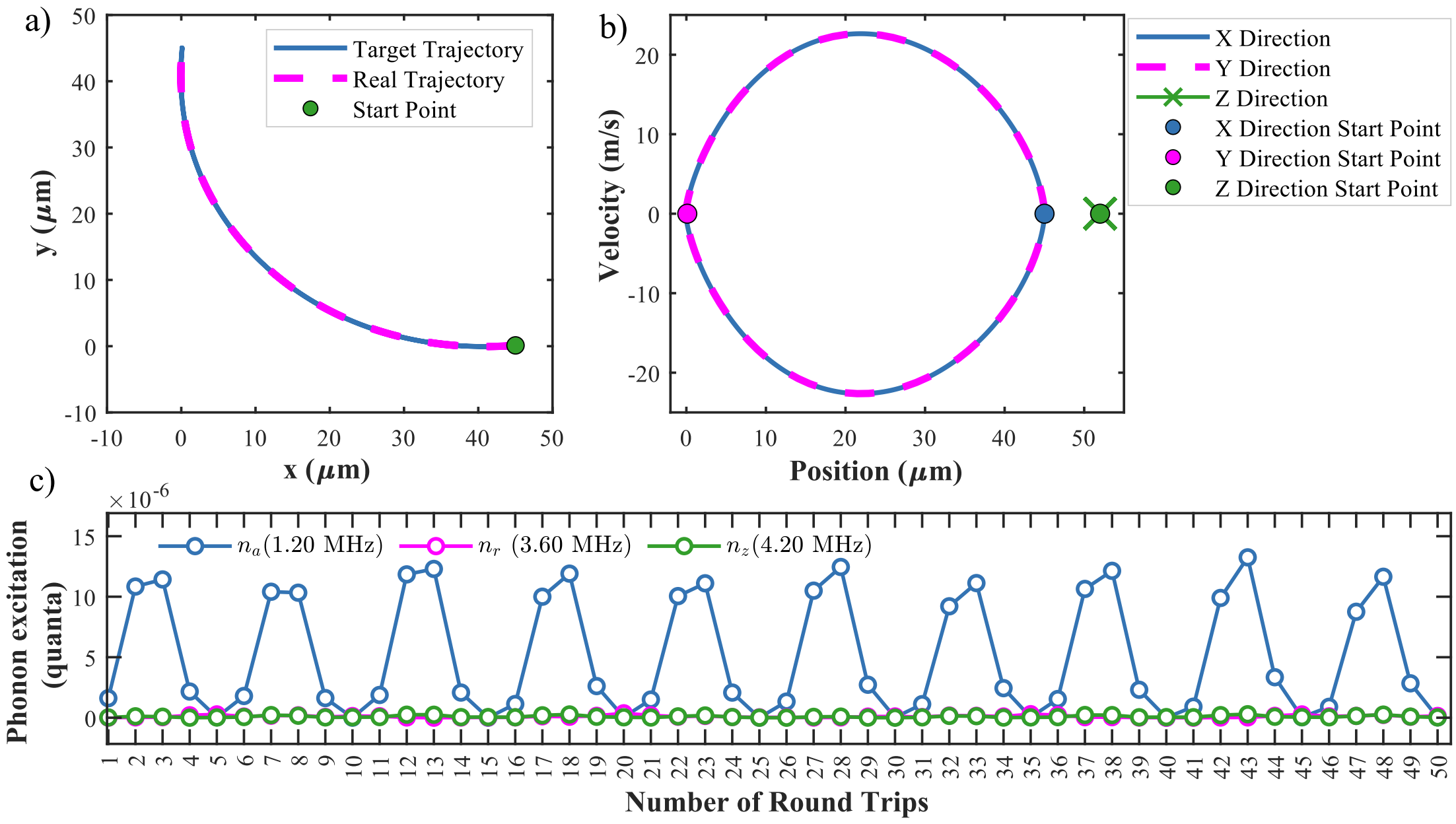

**Fig. 4 | Turning dynamics in the 3D time dependent potential and final coherent excitation after repeated round trips. a**, Target (solid curve) and simulated (dashed curve) trajectories for one round trip through the turn; the green dot marks the starting point. **b**, Position and velocity phase-space trajectories along x, y and z during one round trip; dots mark the initial state in each direction. **c**, Final equivalent coherent phonon numbers of the three local eigenmodes versus the number of consecutive round trips.

**Impact of faster X-junction transport on QCCD execution efficiency**

To assess the system level impact of accelerating individual X-junction operations, we define the transport speedup factor as $S_X = v_X/v_{X,0}$. Here, $v_X$ is the X-junction transport speed and $v_{X,0} = 4$ m/s is the baseline speed, corresponding to $S_X = 1$. The baseline junction crossing time is approximately 187 $\mu s$ (Ref. 17). The average turning speed achieved in this work, approximately 14.6 m/s, corresponds to $S_X \approx 4$ and serves as a representative work point.

Fig. 5a considers the fixed QCCD grid ($G = 5 \times 5$) defined in Ref. 39, comprising 25 grid cells arranged in five rows and five columns and connected by horizontal and vertical transport channels. For an application qubit count $n$, it plots $R_X$, the ratio of X-junction transport time $T_{\text{junc}}$ to total transport time $T_{\text{trans}}$, and its median across six benchmark circuit classes[40–42], against the speedup factor $S_X$. This comparison assesses how accelerating X-junction transport changes the composition of transport time for different circuits. Differences in their responses mainly reflect the relative contributions of the underlying transport operations. At $S_X = 1$, the median $R_X$ across the six circuits is 71.4%, showing that X-junction operations dominate total transport time. At the operating point $S_X = 4$, this fraction decreases to 40.5%, substantially reducing $R_X$.

Fig. 5b-c compares $R_X$ before and after X-junction acceleration across combinations of $G$ and $n$, representing different QCCD computational scales. At $S_X = 1$ (Fig. 5b), the median $R_X$ across the six circuit classes is approximately 62–82%, with overall lower and upper bounds of 57.9% and 92.7%. The fraction generally increases with application size. As transport between zones becomes more frequent, X-junction operations occupy a larger share of transport time and become a major determinant of transport efficiency. Increasing the X-junction speed to $S_X = 4$ reduces the median $R_X$ to approximately 30–62%, with overall bounds of 28.6% and 83.4% (Fig. 5c). Fast X-junction transport therefore substantially reduces its contribution to total transport time.

Reducing $R_X$ also lowers $R_{\text{trans}}$, the fraction of a circuit execution cycle $T_{\text{total}}$ spent on transport $T_{\text{trans}}$. Fig. 5d maps $R_{\text{trans}}$ across G and $n$ at $S_X = 4$. Across combinations of G and $n$, the median $R_{\text{trans}}$ is approximately 54–56%, with overall lower and upper bounds of 48.9% and 57.6%. Relative to $S_X = 1$, acceleration to $S_X = 4$ reduces $R_{\text{trans}}$ from 70% to approximately 55%. Thus, accelerating a single junction crossing operation substantially reduces the contribution of $T_{\text{trans}}$ to total algorithm execution time.

Finally, we assess how reducing $T_{\text{trans}}$ affects $T_{\text{total}}$. Fig. 5e shows $R_T$, the ratio of total execution time at $S_X = 4$ to that at $S_X = 1$. Across the evaluated G and $n$ range, the median $R_T$ across the six circuits is approximately 67%, corresponding to a 33% reduction in total execution time. The overall lower and upper bounds are 58.7% and 70.6%. We further use the system occupancy $\rho = n/2G$ (each grid cell accommodates two computational qubits, the system capacity is 2G) to describe the simultaneous scaling of G and $n$. Four constant occupancy contours, corresponding to $\rho = 20\%$, 40%, 60% and 80%, show that $R_T$ decreases as $\rho$ increases. The smaller $R_T$ at higher occupancy indicates that accelerating individual X-junction operations produces a larger reduction in $T_{\text{total}}$ when more ions participate in computation.

Together, the results in Fig. 5 show that X-junction transport occupies an increasing share of algorithm execution time as two dimensional QCCD systems scale. Increasing its speed is therefore important for improving both circuit execution efficiency and QCCD scalability.

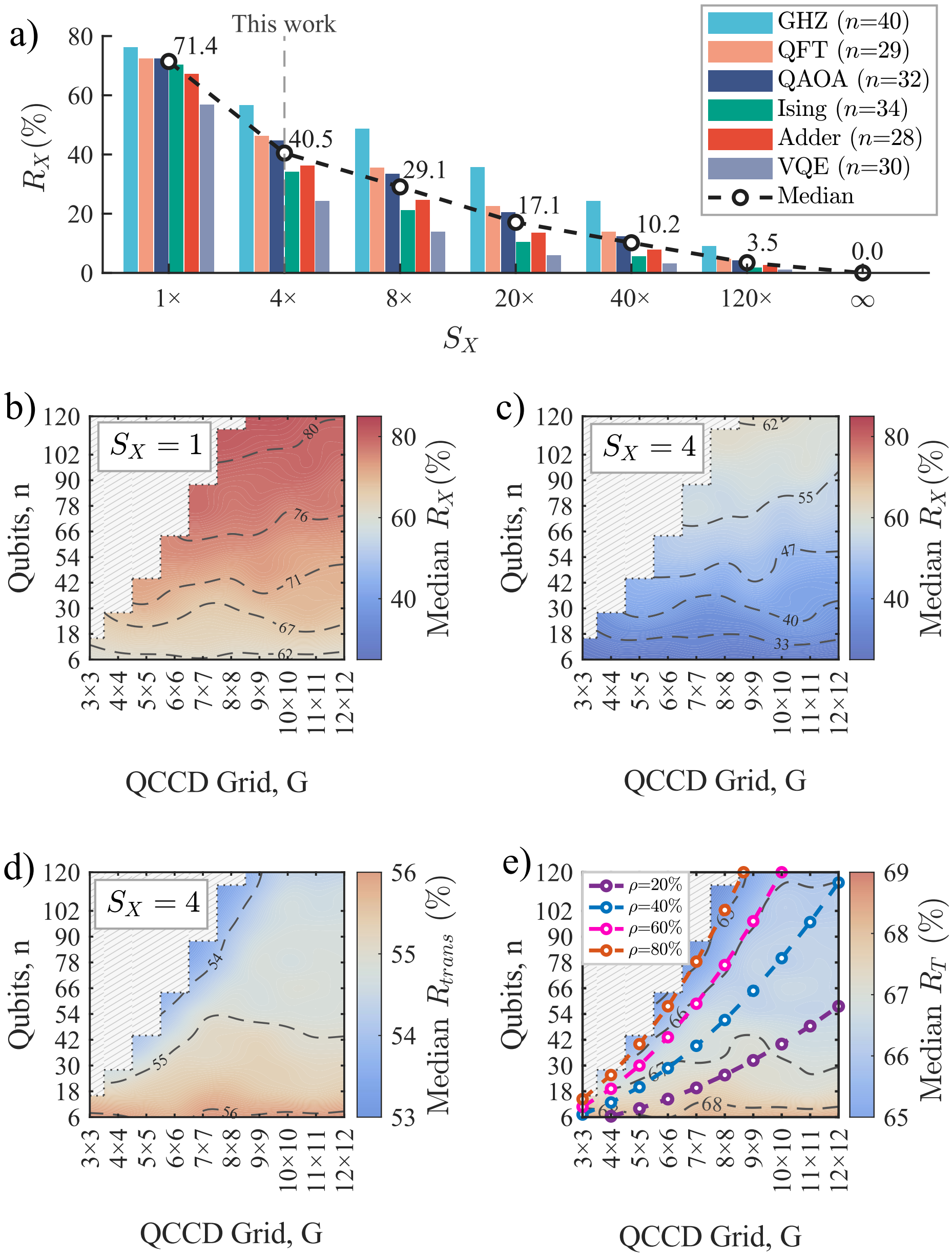

**Fig. 5 | Impact of X-junction acceleration on QCCD transport time and single run circuit execution time.** The baseline assumes that transport accounts for 70% of total time for a single algorithm execution[3]. **a**, Ratio $R_X$ of accelerated X-junction transport time $T_{\text{junc}}$ to total transport time $T_{\text{trans}}$ versus speedup factor $S_X$ for different quantum algorithms. The QCCD grid $G$ is fixed at $5 \times 5$, with application qubit counts $n$ as indicated. **b-c**, Dependence of $R_X$ on G and $n$ before ($S_X = 1$) and after ($S_X = 4$) acceleration, respectively. **d**, Ratio $R_{\text{trans}}$ of $T_{\text{trans}}$ to $T_{\text{total}}$ across G and $n$ at $S_X = 4$. **e**, Ratio $R_T$ of total execution time at $S_X = 4$ to that at $S_X = 1$, across G and $n$. Purple, blue, pink and orange dashed curves indicate constant system occupancies of $\rho = 20\%, 40\%, 60\%$ and $80\%$, respectively. **b–e**, Background colors show medians across the six benchmark circuit classes, and black dashed curves denote contours. Hatched regions mark parameter combinations with $\rho \geq 90\%$.

## Discussion

LHX-T integrates single connected, low barrier pseudopotential channels designed using Multi-RF[24] with multidimensional STA dynamics[22] and coordinated optimization of electrode voltage waveforms. Its main distinction lies in prioritizing dynamical targets and constraints in waveform optimization, rather than merely introducing a new numerical solver. Ion forces and suppression of unwanted motion coupling terms in the Hessian are imposed as equality constraints. Local curvature matching, voltage amplitude regularization, and waveform smoothness are incorporated into a weighted objective. Lagrange multipliers combine these conditions into a KKT system, yielding multichannel voltage waveforms over the entire transport sequence. Three dimensional dynamical simulations show that a single $^{171}Yb^{+}$ ion completes a 90° X-junction turn in 5 $\mu s$ at an average speed of 14.6 m/s. Across 50 consecutive round trips, the final equivalent coherent phonon number of the low frequency mode peaks at only $1.5 \times 10^{-5}$. Its variation is periodic, with no cumulative increase.

These results are based on an idealized single ion numerical model. Stray electric fields, RF amplitude and phase errors, electrode fabrication deviations, and the bandwidth of the control electronics and signal paths are not included. The calculated phonon numbers quantify only residual coherent motion at the endpoint. They exclude initial thermal excitation and additional heating from environmental noise. Experimental measurements and robustness tests are therefore needed to establish practical performance, as well as extensions to multiple ion and mixed species systems.

Performance analysis of QCCD systems with increasing qubit counts indicates that fast transport waveform design can substantially shorten quantum circuit execution cycles. Here, we calculate and demonstrate waveforms only for fast X-junction transport. LHX-T could also serve as a computational tool for designing fast waveforms for linear transport[26], ion chain splitting and merging[43], and rotational swapping[44]. Accelerating these transport primitives could further reduce transport and total computation times in scalable QCCD quantum computers.

## Methods

The LHX-T framework is summarized in Fig. 6. First, the prescribed continuous two dimensional trajectory determines the target force and the local curvature matrix[22]. Multi-RF optimization then constructs a low barrier pseudopotential channel with a nearly constant height along that trajectory[24]. Together with the DC unit voltage basis potentials, this defines a linear mapping from multichannel DC voltages to the target forces and curvatures[15,36]. Essential dynamical conditions are imposed as equality constraints. Curvature matching and voltage smoothness define a quadratic optimization over the entire transport sequence. Lagrange multipliers yield a KKT system for the multichannel voltage sequence[45,46]. Finally, the optimized voltages are substituted into the three dimensional time dependent potential to validate the transport dynamics and the final phonon excitation[37,38].

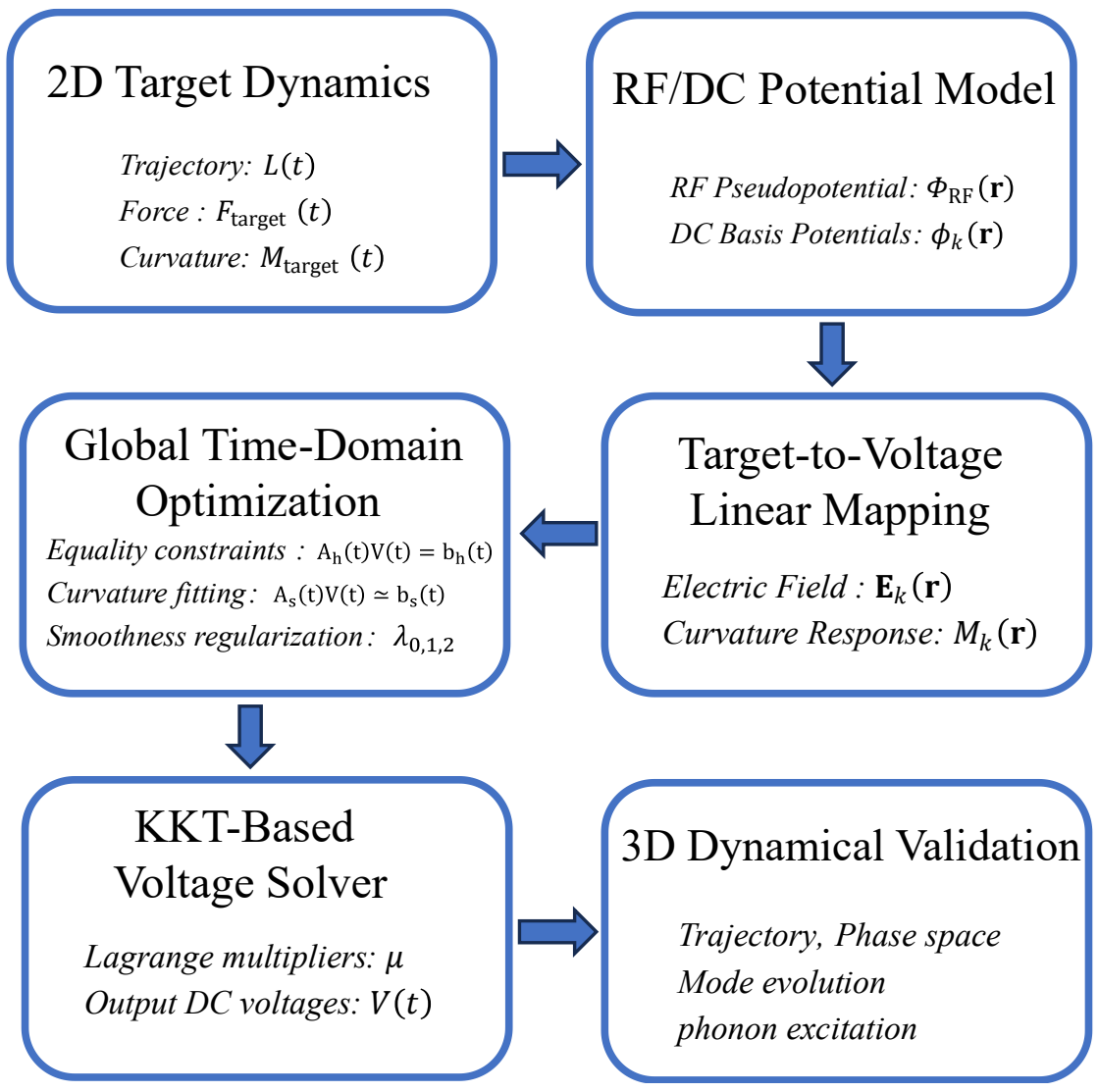

**Fig. 6 | Overview of the LHX-T method.**

### Target dynamics for continuous two dimensional turning

X-junction transport occurs primarily in the $x$–$y$ plane at a fixed height above the chip, so we construct only the in-plane two dimensional STA target dynamics. The DC potential satisfies Laplace's equation, $\partial_{xx}\Phi_{\text{DC}} + \partial_{yy}\Phi_{\text{DC}} + \partial_{zz}\Phi_{\text{DC}} = 0$. Its target curvature along $z$ is therefore determined by the curvatures in the $xy$ plane[47]. The realized $z$ confinement arises jointly from the RF pseudopotential and the DC potential[31]. A local second order expansion near the target trajectory represents the planar potential as a two dimensional time dependent harmonic trap. Its target Hamiltonian[22] is

$$H(t) = \frac{\mathbf{p}^2}{2m} + \frac{m}{2}[\mathbf{X} - \mathbf{C}(t)]^T K_{\text{target}}(t)[\mathbf{X} - \mathbf{C}(t)]. \quad (1)$$

Here, $\mathbf{X} = (x, y)^T$ and $\mathbf{p} = (p_x, p_y)^T$ are the in-plane position and momentum operators, respectively; $m$ is the ion mass. The vector $\mathbf{C}(t)$ denotes the target harmonic trap center, and $K_{\text{target}}(t)$ is a real symmetric matrix whose eigenvalues are the squared angular frequencies of the local modes. In the ideal time dependent harmonic potential, the target trajectory $\mathbf{L}(t)$ and the instantaneous trap center $\mathbf{C}(t)$ satisfy[22]: $\mathbf{C}(t) = \mathbf{L}(t) + K_{\text{target}}^{-1}(t)\ddot{\mathbf{L}}(t)$. The two dimensional target model therefore provides $L(t)$ and $K_{\text{target}}(t)$ as the principal inputs to voltage optimization. These determine the force required for turning transport and the local mode frequencies and orientations, respectively[7,22].

### Three dimensional total potential model

The target force $F_{\text{target}}(t) = m\ddot{\mathbf{L}}(t)$ and $K_{\text{target}}(t) = (q/m)M_{\text{target}}(t)$ must be mapped to drive voltages for the electrodes in Fig. 1, where $q$ is the ion charge and $M_{\text{target}}(t)$ is the target potential Hessian. To this end, we use a three dimensional surface electrode trap model in which the total potential $\Phi_{\text{tot}}(\mathbf{r}, t)$ is a linear superposition of the RF pseudopotential and DC unit voltage basis potentials[36,47]:

$$\Phi_{\text{tot}}(\mathbf{r}, t) = \Phi_{\text{RF}}(\mathbf{r}) + \sum_{k=1}^{N} V_k(t)\phi_k(\mathbf{r}), \quad (2)$$

In Eq. (2), $\boldsymbol{r} = (x, y, z)^T$ is the spatial coordinate. The effective pseudopotential $\Phi_{\text{RF}}(\boldsymbol{r})$ is generated by the segmented RF electrodes driven at 30 MHz with the optimized amplitudes in Fig. 1. The unit voltage basis potential $\phi_k(\boldsymbol{r})$ of the DC electrode $k$ is calculated using CPO (Charged Particle Optics) software[48]. The $V_k(t)$ is the corresponding dynamic DC voltage, and $N$ is the total number of DC electrodes.

### Linear mapping between target dynamics and DC voltages

Once $\Phi_{\text{RF}}(\mathbf{r})$ is fixed, dynamic DC voltages reproduce the target forces and local curvatures at the target trajectory. We collect the DC electrode voltages at time $t$ into the vector $\mathbf{V}(t) = [V_1, (t), V_2, (t) \dots V_N(t)]^T$. Because the DC potential is a linear superposition of unit voltage basis potentials, each electrode's contributions to the local electric field and Hessian depend linearly on $V_k(t)$[47,49]. For electrode $k$, the unit voltage electric field response at $\mathbf{r}$ is $\mathbf{E}_k(\mathbf{r}) = -\nabla\phi_k(\mathbf{r})$, the ion force is $q\mathbf{E}_k(\mathbf{r})$, and the curvature response is $\mathbf{M}_k(\mathbf{r}) = (q/m)\nabla\nabla\phi_k(\mathbf{r})$. The corresponding RF pseudopotential responses are denoted $\mathbf{E}_{\text{RF}}(\mathbf{r})$ and $\mathbf{M}_{\text{RF}}(\mathbf{r})$. At the target trajectory $\mathbf{L}(t)$, the total electric field and local curvature are linear combinations of the electrode voltage contributions:

$$\mathbf{E}_{\text{tot}}[\mathbf{L}(t), t] = \mathbf{E}_{\text{RF}}[\mathbf{L}(t)] + \sum_{k=1}^{N} V_k(t)\mathbf{E}_k[\mathbf{L}(t)], \quad (3)$$

$$M_{\text{real}}[\mathbf{L}(t), t] = M_{\text{RF}}[\mathbf{L}(t)] + \sum_{k=1}^{N} V_k(t)M_k[\mathbf{L}(t)]. \quad (4)$$

The RF pseudopotential is fixed by the RF electrode geometry and drive voltages, whereas DC optimization adjusts only the dynamic DC potential. The RF contributions must therefore be subtracted from the target constraints. We divide these target constraints into two classes. Local trap curvature $M_{\text{target}}(t)$ is matched by least squares using its three independent components, $xx$, $yy$ and $xy$:

$$A_s(t) = \begin{bmatrix} M_{1,xx} & M_{2,xx} & \cdots & M_{N,xx} \\ M_{1,yy} & M_{2,yy} & \cdots & M_{N,yy} \\ M_{1,xy} & M_{2,xy} & \cdots & M_{N,xy} \end{bmatrix}_{\mathbf{r}=\mathbf{L}(t)}, \quad (5)$$

$$b_s(t) = \begin{bmatrix} M_{\text{target},xx}(t) - M_{\text{RF},xx} \\ M_{\text{target},yy}(t) - M_{\text{RF},yy} \\ M_{\text{target},xy}(t) - M_{\text{RF},xy} \end{bmatrix}_{\mathbf{r}=\mathbf{L}(t)}. \quad (6)$$

The curvature target in the $xy$ plane is written as $A_s(t)V(t) \simeq b_s(t)$. The total potential Hessian components $M_{xx}$, $M_{yy}$ and $M_{xy}$ jointly determine the local in-plane mode orientations. The second class comprises constraints that must be satisfied: the ion force, force balance along $z$, and suppression of cross coupling terms $M_{xz}$ and $M_{yz}$. These are expressed as explicit equality constraints, $A_h(t)V(t) = b_h(t)$, where $A_h(t)$ contains the corresponding linear electrode responses:

$$A_h(t) = \begin{bmatrix} qE_{1x} & qE_{2x} & \cdots & qE_{Nx} \\ qE_{1y} & qE_{2y} & \cdots & qE_{Ny} \\ qE_{1z} & qE_{2z} & \cdots & qE_{Nz} \\ M_{1,xz} & M_{2,xz} & \cdots & M_{N,xz} \\ M_{1,yz} & M_{2,yz} & \cdots & M_{N,yz} \end{bmatrix}_{\mathbf{r}=\mathbf{L}(t)}, \quad (7)$$

$$b_h(t) = \begin{bmatrix} m\ddot{L}_x(t) - qE_{\mathrm{RF},x} \\ m\ddot{L}_y(t) - qE_{\mathrm{RF},y} \\ -qE_{\mathrm{RF},z} \\ -M_{\mathrm{RF},xz} \\ -M_{\mathrm{RF},yz} \end{bmatrix}_{\mathbf{r}=\mathbf{L}(t)}. \quad (8)$$

The first three rows of Eq. (8) impose the target forces along $x, y$ and force balance along $z$. Fixed height transport requires $\ddot{L}_z(t) = 0$. The final two rows suppress $M_{xz}$ and $M_{yz}$, thereby suppress coupling between the $xy$ modes and the z mode[31]. At each time point, voltage mapping thus combines a curvature matching objective that permits trade offs with linear equality constraints that must be satisfied. Under finite electrode degrees of freedom, this formulation satisfies the equalities while retaining scope to optimize voltage amplitudes and temporal smoothness.

**Constrained voltage optimization over the entire transport sequence**

To ensure temporal continuity across channels, we combine the voltages at all discrete time points and DC channels into a single optimization variable $\boldsymbol{V}$:

$$V = \left[V^T(t_1), V^T(t_2), \dots, V^T(t_Q)\right]^T, \quad (9)$$

Here, $Q = 1000$ is the number of discrete time points and $V(t_i)$ is the DC voltage vector at time point $i$. The curvature fitting matrices and target vectors are assembled into $A_s$ and $b_s$, while the explicit equality constraints are assembled into $A_h$ and $b_h$. Simultaneous optimization of voltage amplitudes and first and second order temporal differences[37,50,51] yields smooth, continuous waveforms. The optimization over the entire transport sequence is written as

$$f(V) = \min_V \frac{1}{2}\|A_s V - b_s\|^2 + \frac{\lambda_0}{2}\|V - V_{ref}\|^2 + \frac{\lambda_1}{2}\|\mathrm{D}V\|^2 + \frac{\lambda_2}{2}\|\mathrm{D}^2 V\|^2, \quad \text{s.t.} \quad A_h V = b_h. \quad (10)$$

The first term in Eq. (10) fits local trap curvature by least squares. The second penalizes excessive voltage amplitudes with reference voltage $V_{ref} = 10V$. The third penalizes the rate of voltage change using the first order temporal difference operator D, thereby providing slew rate control. The fourth uses the second order operator $D^2$ to penalize rapid changes in the voltage slope, suppressing high frequency oscillations and improving waveform smoothness. The parameters $\lambda_0 = 10^{-2}, \lambda_1 = 10^3$ and $\lambda_2 = 10^7$ are the regularization weights for voltage amplitude, slew rate, and waveform smoothness, respectively.

Expanding Eq. (10) and collecting its quadratic and linear terms gives the standard form of an equality constrained quadratic optimization problem[45,46]:

$$f(V) = \min_V \frac{1}{2} V^T P V + g^T V, \quad \text{s.t.} \quad A_h V = b_h, \quad (11)$$

Where

$$P = A_s^T A_s + \lambda_0 I + \lambda_1 D_1^T D_1 + \lambda_2 D_2^T D_2,$$
$$g = -A_s^T b_s - \lambda_0 V_{ref}. \quad (12)$$

Introducing Lagrange multipliers $\mu$, we construct the corresponding Lagrangian as follows:

$$\mathcal{L}(V, \mu) = \frac{1}{2} V^T P V + g^T V + \mu^T (A_h V - b_h). \quad (13)$$

The first order optimality conditions $\frac{\partial \mathcal{L}}{\partial V} = 0, \frac{\partial \mathcal{L}}{\partial \mu} = 0$ give the following pair of equations:

$$PV + g + A_h^T \mu = 0, \qquad A_h V = b_h. \quad (14)$$

These first-order optimality conditions yield the following KKT linear system:

$$\begin{bmatrix} P & A_h^T \\ A_h & 0 \end{bmatrix} \begin{bmatrix} V \\ \mu \end{bmatrix} = \begin{bmatrix} -g \\ b_h \end{bmatrix}. \quad (15)$$

Solving Eq. (15) gives the multichannel DC voltage sequence $V$ over the entire transport interval. The formulation imposes explicit equality constraints on the target forces, the z force balance, and the suppression of unwanted Hessian cross coupling terms. Local curvature matching, voltage amplitude regularization, and temporal smoothness remain in the quadratic objective. Smooth waveforms are thus determined within the feasible space that prioritizes the essential dynamical constraints.

**Numerical dynamics and final phonon excitation**

We simulate a single $^{171}Yb^+$ ion along the target trajectory in Fig. 1 at a height of 52 $\mu m$. The dynamics are obtained by numerically integrating Newton's equations in the realized three dimensional, time dependent total potential $\Phi_{\mathrm{tot}}(\mathbf{r}, t)$. Position and velocity are calculated with MATLAB[52] using ode45, with RelTol = $10^{-12}$, AbsTol = $10^{-13}$, and MaxStep = $10^{-8}$ s. For each round trip, the waveforms in Fig. 2a–d drive the forward turn, and their time reversed sequence drives the return. This procedure is repeated for 50 consecutive round trips flights. At each round trip endpoint, the residual displacement $\Delta s_i$ from the final target local trap center and the residual velocity $\Delta v_i$ are projected onto the local eigenmode $i$. The equivalent coherent phonon number is calculated as $m(\Delta v_i{}^2 + \omega_i{}^2 \Delta s_i{}^2)/(2\hbar\omega_i)$, where $\omega_i$ is the final local mode frequency and $\hbar$ is the reduced Planck constant. This quantity measures only the coherent motional energy associated with residual phase-space displacement. It excludes zero point energy, initial thermal occupation, and environmental heating[32,38].

**Assessing the impact of fast X-junction transport on QCCD system performance**

Table 1 summarizes experimentally reported times for the main QCCD transport operations. We use the transport, cooling, and quantum gate operation times from Ref. 3,12,53 as the baseline. These values determine the total ion transport time $T_{\mathrm{trans}}$ and total time $T_{\mathrm{total}}$ for a single algorithm execution across different two dimensional QCCD grid sizes. Following Ref. 3, we set the baseline ratio $T_{\mathrm{trans},0}/T_{\mathrm{total},0}$ to 70%. Only the X-junction transport time is varied. Gate, cooling and other non-transport operation times within each computation cycle remain fixed, isolating the impact of X-junction acceleration on overall runtime. The assessment uses the S-SYNC framework[25]. Six representative circuit classes, Adder, GHZ, Ising, QFT, QAO and VQE, are selected from an open source benchmark suite[40–42] to reduce dependence on any single circuit structure.

**Table 1 | Experimental times for basic QCCD transport operations and values used in this work**

| Operation | Reported time range ($\mu s$) | Time used in this work ($\mu s$) |
|---|---|---|
| Linear transport[4,36,37,53–56] | 3.6 – 58 | 6 |
| Splitting/merging[4,38,58,59] | 55 – 375 | 80 |
| X-junction transport[16,17,57,60] | 140 – 200 | 187 |
| Reordering[4,61,62] | 25 – 50 | 42 |

**Data availability**

The data corresponding to this study is available upon reasonable request.

**Code availability**

The code corresponding to this study is available upon reasonable request.

## References

1. Bruzewicz, C. D., Chiaverini, J., McConnell, R. & Sage, J. M. Trapped-ion quantum computing: Progress and challenges. *Applied Physics Reviews* **6**, 021314 (2019).
2. Haffner, H., Roos, C. & Blatt, R. Quantum computing with trapped ions. *Physics Reports* **469**, 155–203 (2008).
3. Ransford, A. *et al.* A 98-qubit trapped-ion quantum computer with all-to-all connectivity. *Nature* **655**, 81–86 (2026).
4. Pino, J. M. *et al.* Demonstration of the trapped-ion quantum CCD computer architecture. *Nature* **592**, 209–213 (2021).
5. Brown, K. R., Kim, J. & Monroe, C. Co-designing a scalable quantum computer with trapped atomic ions. *npj Quantum Inf* **2**, 16034 (2016).
6. Enzer, D. G. *et al.* Observation of power-law scaling for phase transitions in linear trapped ion crystals. *Phys. Rev. Lett.* **85**, 2466–2469 (2000).
7. James, D. F. V. Quantum dynamics of cold trapped ions with application to quantum computation. *Appl. Phys. B: Lasers Opt.* **66**, 181–190 (1998).
8. Landsman, K. A. *et al.* Two-qubit entangling gates within arbitrarily long chains of trapped ions. *Phys. Rev. A* **100**, 022332 (2019).
9. Kielpinski, D., Monroe, C. & Wineland, D. J. Architecture for a large-scale ion-trap quantum computer. *Nature* **417**, 709–711 (2002).
10. Home, J. P. *et al.* Complete methods set for scalable ion trap quantum information processing. *Science* **325**, 1227–1230 (2009).
11. Kaushal, V. *et al.* Shuttling-based trapped-ion quantum information processing. *AVS Quantum Science* **2**, 014101 (2020).
12. Moses, S. A. *et al.* A race-track trapped-ion quantum processor. *Phys. Rev. X* **13**, 041052 (2023).
13. Hensinger, W. K. *et al.* T-junction ion trap array for two-dimensional ion shuttling, storage, and manipulation. *Appl. Phys. Lett.* **88**, 034101 (2006).
14. Blakestad, R. B. *et al.* High-fidelity transport of trapped-ion qubits through an X-junction trap array. *Phys. Rev. Lett.* **102**, 153002 (2009).
15. Blakestad, R. B. *et al.* Near-ground-state transport of trapped-ion qubits through a multidimensional array. *Phys. Rev. A* **84**, 032314 (2011).
16. Wright, K. *et al.* Reliable transport through a microfabricated *X*-junction surface-electrode ion trap. *New J. Phys.* **15**, 033004 (2013).
17. Burton, W. C. *et al.* Transport of Multispecies Ion Crystals through a Junction in a Radio-Frequency Paul Trap. *Phys. Rev. Lett.* **130**, 173202 (2023).
18. Dai, W., Brown, K. A. & Robertazzi, T. G. Advanced shuttle strategies for parallel QCCD architectures. *IEEE Trans. Quantum Eng.* **5**, 1–18 (2024).
19. Webber, M., Herbert, S., Weidt, S. & Hensinger, W. K. Efficient qubit routing for a globally connected trapped ion quantum computer. *Adv Quantum Tech* **3**, 2000027 (2020).
20. Moehring, D. L. *et al.* Design, fabrication and experimental demonstration of junction surface ion traps. *New J. Phys.* **13**, 075018 (2011).
21. Conta, A., Bogino, S., Köhncke, F., Schmidt-Kaler, F. & Poschinger, U. G. Toolchain for shuttling trapped-ion qubits in segmented traps. *AVS Quantum Science* **8**, 023801 (2026).
22. Simsek, S. & Mintert, F. Quantum control with a multi-dimensional Gaussian quantum invariant. *Quantum* **5**, 409 (2021).
23. Singer, K. *et al.* Colloquium: Trapped ions as quantum bits: Essential numerical tools. *Rev. Mod. Phys.* **82**, 2609–2632 (2010).
24. Liu, Y. *et al.* Cooperative engineering the multiple radio-frequency fields to reduce the X-junction barrier for ion trap chips. *Chip* **3**, 100078 (2024).
25. Zhu, C., Wu, X., Wang, J. & Wang, X. S-SYNC: Shuttle and swap co-optimization in quantum charge-coupled devices. *Proceedings of The 52nd Annual International Symposium On Computer Architecture* 271–284 (2025) doi:10.1145/3695053.3731084.
26. Torrontegui, E. *et al.* Fast atomic transport without vibrational heating. *Phys. Rev. A* **83**, 013415 (2011).
27. Guéry-Odelin, D. *et al.* Shortcuts to adiabaticity: Concepts, methods, and applications. *Rev. Modern Phys.* **91**, 045001 (2019).
28. Lu, X.-J., Lizuain, I & Muga, J. G. Inverse engineering of fast state transfer among coupled oscillators. *Quantum* **6**, 740 (2022).
29. Tobalina, A., Torrontegui, E., Lizuain, I., Palmero, M. & Muga, J. G. Invariant-based inverse engineering of time-dependent, coupled harmonic oscillators. *Phys. Rev. A* **102**, 063112 (2020).
30. Ruzic, B. P. *et al.* Entangling-gate error from coherently displaced motional modes of trapped ions. *Phys. Rev. A* **105**, 052409 (2022).
31. Zhang, C., Mehta, K. K. & Home, J. P. Optimization and implementation of a surface-electrode ion trap junction. *New J. Phys.* **24**, 073030 (2022).
32. Leibfried, D., Blatt, R., Monroe, C. & Wineland, D. Quantum dynamics of single trapped ions. *Rev. Mod. Phys.* **75**, 281 (2003).
33. Saito, R., Saito, K. & Mukaiyama, T. Measurement of ion displacement via RF power variation for excess micromotion compensation. *Journal of Applied Physics* **129**, 124302 (2021).
34. Davis, C. The rotation of eigenvectors by a perturbation. *Journal of Mathematical Analysis and Applications* **6**, 159–173 (1963).
35. Fabrikant, M. I., Lauria, P., Madjarov, I. S., Burton, W. C. & Sutherland, R. T. Cooling trapped ions with phonon rapid adiabatic passage. *Phys. Rev. X* **14**, 041046 (2024).
36. House, M. G. Analytic model for electrostatic fields in surface-electrode ion traps. *Phys. Rev. A* **78**, 033402 (2008).
37. Walther, A. *et al.* Controlling fast transport of cold trapped ions. *Phys. Rev. Lett.* **109**, 080501 (2012).
38. Bowler, R. *et al.* Coherent diabatic ion transport and separation in a multizone trap array. *Phys. Rev. Lett.* **109**, 080502 (2012).
39. Wu, X., Zhu, C., Wang, J. & Wang, X. MUSS-TI: multi-level shuttle scheduling for large-scale entanglement module linked trapped-ion. in 749–763 (ACM, Seoul Korea, 2025). doi:10.1145/3725843.3756129.
40. Li, A., Stein, S., Krishnamoorthy, S. & Ang, J. QASMBench: A low-level quantum benchmark suite for NISQ evaluation and simulation. *ACM Transactions on Quantum Computing* **4**, 1–26 (2023).
41. Kandala, A. *et al.* Hardware-efficient variational quantum eigensolver for small molecules and quantum magnets. *Nature* **549**, 242–246 (2017).
42. Moll, N. *et al.* Quantum optimization using variational algorithms on near-term quantum devices. *Quantum Sci. Technol.* **3**, 030503 (2018).
43. Palmero, M., Martínez-Garaot, S., Poschinger, U. G., Ruschhaupt, A. & Muga, J. G. Fast separation of two trapped ions. *New J. Phys.* **17**, 093031 (2015).
44. Tobalina, A., Muga, J. G., Lizuain, I. & Palmero, M. Shortcuts to adiabatic rotation of a two-ion chain. *Quantum Sci. Technol.* **6**, 045023 (2021).
45. Boyd, S. P. & Vandenberghe, L. *Convex Optimization*. (Cambridge University Press, Cambridge, UK; New York, 2004).
46. Gould, N. I. M., Hribar, M. E. & Nocedal, J. On the solution of equality constrained quadratic programming problems arising in optimization. *SIAM J. Sci. Comput.* **23**, 1376–1395 (2001).
47. Wesenberg, J. H. Electrostatics of surface-electrode ion traps. *Phys. Rev. A* **78**, 063410 (2008).

48. Read, F. H. & Bowring, N. J. The CPO programs and the BEM for charged particle optics. *Nuclear Instruments and Methods in Physics Research Section A: Accelerators, Spectrometers, Detectors and Associated Equipment* **645**, 273–277 (2011).
49. Hucul, D. *et al.* On the transport of atomic ions in linear and multidimensional ion trap arrays. *QIC* **8**, 501–578 (2008).
50. Orozco-Ruiz, M. *et al.* Optimal control with a multidimensional quantum invariant. *Phys. Rev. A* **108**, 022601 (2023).
51. Reichle, R. *et al.* Transport dynamics of single ions in segmented microstructured Paul trap arrays. *Fortschritte der Physik* **54**, 666–685 (2006).
52. The MathWorks Inc. (2023). MATLAB version: 9.14.0.2206163 (R2023a), Natick, Massachusetts: The MathWorks Inc.
53. Wu Y.-K. & Duan L.-M. Research progress of ion trap quantum computing. *Acta Phys. Sin.* **72**, 230302 (2023).
54. Huber, G. *et al.* Transport of ions in a segmented linear paul trap in printed-circuit-board technology. *New J. Phys.* **10**, 013004 (2008).
55. Palmero, M., Bowler, R., Gaebler, J. P., Leibfried, D. & Muga, J. G. Fast transport of mixed-species ion chains within a paul trap. *Phys. Rev. A* **90**, 053408 (2014).
56. Sterk, J. D. *et al.* Closed-loop optimization of fast trapped-ion shuttling with sub-quanta excitation. *npj Quantum Inf* **8**, 68 (2022).
57. Delaney, R. D. *et al.* Scalable multispecies ion transport in a grid-based surface-electrode trap. *Phys. Rev. X* **14**, 041028 (2024).
58. Ruster, T. *et al.* Experimental realization of fast ion separation in segmented paul traps. *Phys. Rev. A* **90**, 033410 (2014).
59. Lancellotti, F. *et al.* Low-excitation transport and separation of high-mass-ratio mixed-species ion chains. *Phys. Rev. Research* **6**, L032059 (2024).
60. Blakestad, R. B. Transport of Trapped-Ion Qubits within a Scalable Quantum Processor. (Colorado\NIST, 2010).
61. Kaufmann, H. *et al.* Fast ion swapping for quantum-information processing. *Phys. Rev. A* **95**, 052319 (2017).
62. Van Mourik, M. W. *et al.* Coherent rotations of qubits within a surface ion-trap quantum computer. *Phys. Rev. A* **102**, 022611 (2020).

## Acknowledgements
This work was supported by the Funding of Joint Laboratory of Quantum Computing Integration Technology, MoonQuest AI & GDUT (26HK0336).

## Author contributions
J.X. and Z.W. contributed equally to this work and jointly conceived the study, designed the overall research framework, and developed the theoretical model and methodology. J.X. and C.Z. performed the main numerical calculations and conducted the data analysis and visualization. J.X. and S.G. drafted the initial manuscript. C.Z. and X.W. contributed part of the data presented in the Results and provided related technical support. F.D. and X.W. supervised the project and provided guidance on the research design and interpretation of the results. All authors discussed the results, revised the manuscript, and approved the final version.

## Competing interests
The authors declare no competing interests.

## Additional information
Correspondence and requests for materials should be addressed to Xu Wang.